\documentclass{jps-cp}
\usepackage{txfonts} 
\usepackage{siunitx}

\title{Development of a Neutron Imager Using CMOS Pixelated Sensor}

\author{
Yoshio \textsc{Kamiya}$^{1}$, Hidetoshi \textsc{Ohshita}$^{2}$, Ryutaro \textsc{Nishimura}$^{2,3}$, Toshiya \textsc{Muto}$^{4}$,
Shingo \textsc{Mitsui}$^{5}$, Tomohiro \textsc{Seya}$^{2}$,  Izumi \textsc{Umegaki}$^{2,3}$, and Yasuo \textsc{Arai}$^{6}$
}

\inst{
$^{1}$Department of Physics and International Center for Elementary Particle Physics, The University of Tokyo, Tokyo 113-0033, Japan \\
$^{2}$Institute of Materials Structure Science, High Energy Accelerator Research Organization (KEK), Ibaraki 305-0801, Japan \\
$^{3}$The Graduate University for Advanced Studies (SOKENDAI), Kanagawa 240-0193, Japan \\
$^{4}$The Research Center for Accelerator and Radioisotope Science (RARiS), Tohoku University, Miyagi 982-0826, Japan \\
$^{5}$Data Science and AI Innovation Research Promotion Center, Shiga University, Shiga 522-8522, Japan \\
$^{6}$Accelerator Laboratory, High Energy Accelerator Research Organization (KEK), Ibaraki 305-0801, Japan \\
}

\email{kamiya@icepp.s.u-tokyo.ac.jp}
\recdate{February 28, 2025}

\abst{
We acquired an image of the Siemens Star Chart using the ${}^{10}$B-INTPIX4 CMOS pixelated imager at J-PARC MLF BL21 (NOVA).
The image blurriness, characterized by a root mean square (rms) value of 17 $\mu$m in the line spread function, 
corresponds to an rms beam divergence of 2.7 mrad. for the neutron beam.
This is equivalent to the effective geometrical ratio (L/D ratio) of 370.
The contrast of the image was analyzed as a function of structural scale of the chart,
and based on this contrast, the modulation transfer function (MTF) of the imaging system was evaluated.
The results indicate that the structure corresponding to the 50\% cutoff is 12 line-space pairs per mm.
}

\kword{neutron imaging, monolithic CMOS sensor, spacial resolution, modulation transfer function}

\begin{document}
\maketitle

\section{Introduction}

With the recent development of high-intensity neutron sources, neutron-based materials research and industrial applications are becoming increasingly important. 
Just as the establishment of synchrotron radiation facilities has driven industrial progress, 
advancements in neutron facilities and enhancements to their user environment are expected to yield significant societal benefits.
To expand the potential applications of neutron facilities, 
we have developed a high-precision neutron imager based on a CMOS pixel sensor. 
When combined with a thin conversion material, such as boron or lithium, which converts neutrons into charged particles, 
this imager shows promise for achieving a resolution below 5 $\mu$m in neutron detection \cite{ref:b10int4}. 

In this paper, we use the SOIPIX-INTPIX4 monolithic CMOS pixel sensor \cite{ref:int4}
and present the results of imaging a Siemens Star Chart, a standard tool for evaluating imaging performance, 
at the J-PARC, Materials and Life Science Experimental Facility (MLF) -- Beam Line 21 (BL21/NOVA) \cite{ref:nova}.
To understand the capabilities of the high-precision imaging system, we evaluated the image blurring using multiple metrics, including the root mean square (rms) in the line spread function (LSF),
the modulation transfer function (MTF), the 50\% cutoff line-space pair width, and the effective geometrical ratio (L/D ratio).

The structure of this paper is as follows: 
Section 2 introduces the experimental components, 
Section 3 discusses the selection of neutron events, 
and Section 4 examines the imaging accuracy. 
Finally, Section 5 presents the summary.

\section{Experimental Components}

SOIPIX \cite{ref:soi} is a sensor family fabricated using the 0.2 $\mu$m CMOS 
FD-SOI process \cite{ref:oki} by OKI Semiconductor Co. Ltd. (now LAPIS Semiconductor Co. Ltd.). 
This process enables the integration of the front-end circuit on a semiconductor layer above an insulating layer, 
allowing for a monolithic structure that eliminates the need for a bump bonding process with the sensing nodes. 
Its simple architecture is expected to reduce parasitic capacitance, facilitating high-speed operation of the front-end circuit. 
Additionally, discussions suggest that this design could help lower manufacturing costs in mass production. 
The use of high-resistivity silicon crystals on the backside of the insulating layer 
helps maintain a thick sensitive layer with a deep depletion region, 
making it suitable for detecting highly penetrating particles such as muons.

INTPIX4, a member of this series, features a charge integration circuit in its front end. 
It has been developed as a versatile sensor for a wide range of applications. 
The basic specifications of the sensor are as follows: 
the pixel size is $17 \times 17$ $\mu$m$^{2}$, with an active area of $14.1 \times 8.7$ mm$^{2}$ ($832 \times 512$ pixels),
the readout time is 280 ns per pixel, and the wafer thickness is 300 $\mu$m.\\

The signal readout is performed using the PF-DAQSIX system, 
which has been developed at the Photon Factory (PF) in KEK.
For more detailed information, please refer to reference \cite{ref:DAQSIX}. 
The Kintex-7 FPGA on the DAQ main board generates the control signals for the sensor, 
which are distributed via a relay and distribution board to accommodate up to 4 sensors. 
Signals from the sensors are amplified, digitized using a 12-bit ADC, and transmitted to the data aquisition computer. 
A high-speed 10 Gbps Ethernet interface is integrated into the main board.\\

A enriched boron, ${}^{10}$B, is used to convert neutrons into charged particles,
via $^{10}\mathrm{B}\,(n, \alpha){}^{7}\mathrm{Li}$ interactions:
\begin{eqnarray*}
{}_{(93.9\%)} ~~~ n + {}^{10}{\rm B} \!\!\!&\rightarrow&\!\!\! \alpha_{(1.47 ~{\rm MeV})} + {}^{7}{\rm Li}_{(0.84 ~{\rm MeV})} + \gamma_{(0.48 ~{\rm MeV})} ~,\\
{}_{(6.1\%)} ~~~ n + {}^{10}{\rm B} \!\!\!&\rightarrow&\!\!\! \alpha_{(1.78 ~{\rm MeV})} + {}^{7}{\rm Li}_{(1.01 ~{\rm MeV})} ~.
\end{eqnarray*}
The conversion layer is deposited on the backside of the INTPIX4 sensor 
using Ar sputtering technology, with a thickness of approximately 200 nm.
This imager configuration is referred to as ${}^{10}$B-INTPIX4, hereafter.

Previously, an imaging experiment was conducted at J-PARC MLF BL10 (NOBORU) \cite{ref:noboru}, 
where a Gd-contained line-space pattern was captured and analyzed.
The results confirmed that the spatial resolution, as evaluated by the rms of the LSF, was below 5 $\mu$m \cite{ref:b10int4}.
More recently, simultaneous detection of charged particle pairs from the converter had been demonstrated by sandwiching it between two sensors \cite{ref:b10int4sw}.
The sensor's response to neutrons can be regarded well understood.\\

\begin{figure}[thb]
\begin{center}
\includegraphics[width=0.85\linewidth]{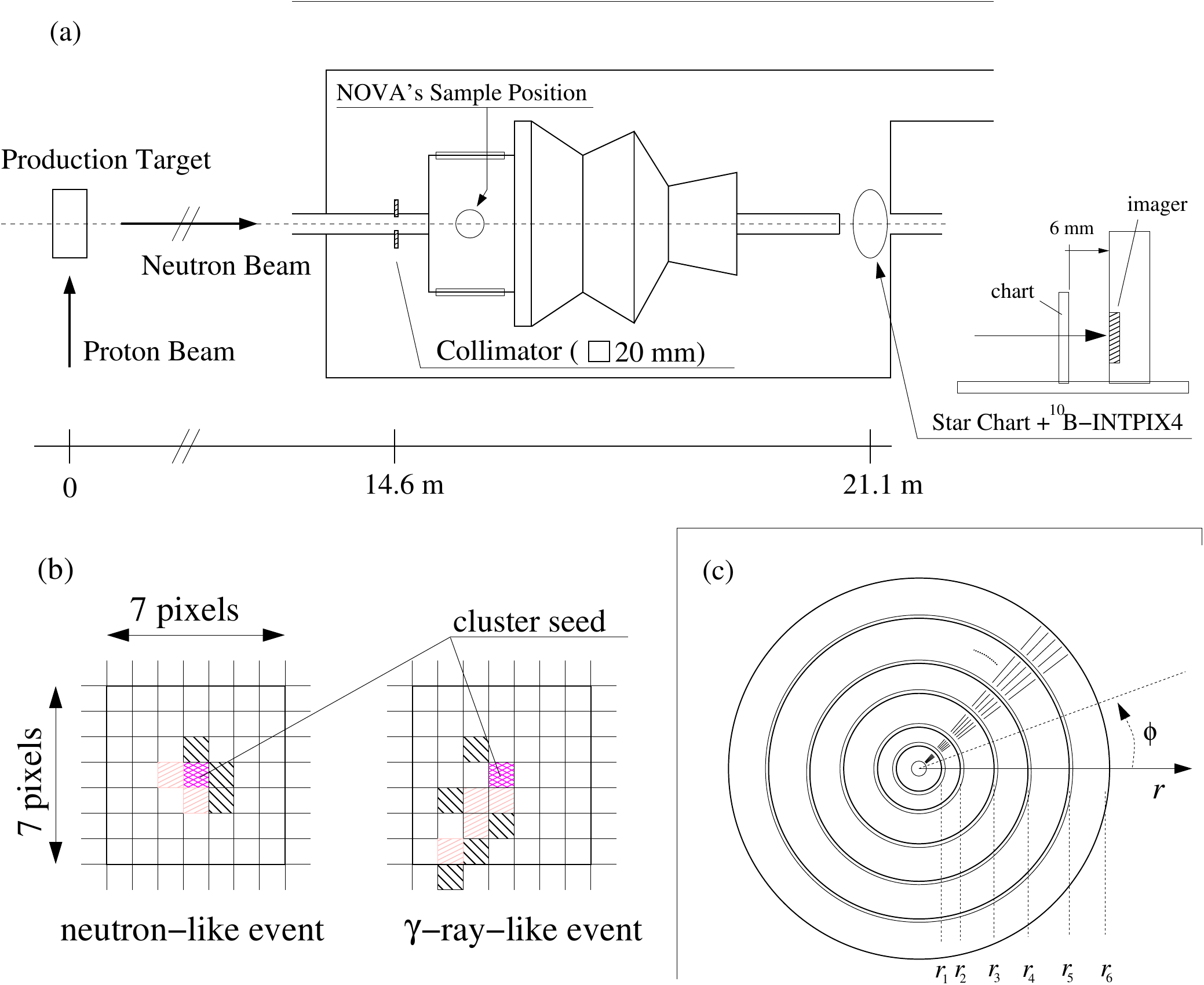}
\caption{
(a) Schematic view of NOVA beamline.
${}^{10}$B-INTPIX4 was installed 21 meters downstream from the neutron production target with a moderator, 
just before the beam dump duct. 
(b) Typical cluster shape for neutron-like/$\gamma$-ray-like event.
(c) Schematic of a Siemens Star Chart.
The Gd pattern is etched in a doughnut-shaped structure.
Each ring contains 128 line-space pairs per full rotation. 
}
\label{fig:nova}
\end{center}
\end{figure}

The experiment with the Siemens Star Chart was conducted using BL21 (NOVA) at J-PARC MLF \cite{ref:nova}.
Figure \ref{fig:nova}(a) provides an overview of the setup.
NOVA is a total diffractometer equipped with arrays of ${}^3$He filled position sensitive detectors covering a large solid angle. 
It enables structural analysis of various materials, including both crystalline and amorphous substances, across a broad Q-range.

${}^{10}$B-INTPIX4 was placed 6.1 m downstream from the NOVA's standard sample location, 
at a distance of 21 m from the neutron production target with a moderator, just before the neutron beam dump duct. 
The neutrons primarily travel through a vacuum duct, 
with the exception of sections totalling approximately 3.5 m (3.1 m upstream and 0.4 m downstream of the sample chamber) where they pass through air.
The neutron beam provided covers a wavelength range from 0.12 {\AA} to 8.3 {\AA} and can be considered to behave as a quantum beam for our device. 
On the sensor surface, the beam had an approximate size of more than $40 \times 40$ mm$^2$, illuminating the entire sensor area. 
A 20 mm $\times$ 20 mm collimator is located 6.5 meters upstream of the imager, determining the geometrical L/D ratio of the beamline to be 330.\\

The Siemens Star Chart is made from a 5-$\mu$m-thick Ge film 
with partially etched regions, featuring radial symmetry divided into 128 segments. 
Figure 1(c) shows a schematic of the star chart, with a repeating angle defined as $d\phi = 2\pi/128$ rad.
The pattern is structured into six concentric donut-shaped regions, all sharing the same center. 
The outer radii of the doughnut-shaped pattern, measured in pixel units, are as follows:
$r_1 = 60$ pixels, 
$r_2 = 120$ pixels,
$r_3 = 240$ pixels,
$r_4 = 361$ pixels, 
$r_5 = 481$ pixels,
$r_6 = 601$ pixels.
These correspond to line-space pair pitches of 
$p_1 = 50 ~\mathrm{\mu m}$, 
$p_2 = 100  ~\mathrm{\mu m}$, 
$p_3 = 200  ~\mathrm{\mu m}$, 
$p_4 = 300  ~\mathrm{\mu m}$, 
$p_5 =  400 ~\mathrm{\mu m}$,
$p_6 =  500 ~\mathrm{\mu m}$, respectively.
It was located 6 mm upstream from the $^{10}$B-INTPIX4 sensor surface.
This configuration corresponds to a 6-$\mu$m blur in the measured image,
when the neutron beam has an rms divergence of 1 mrad.
\begin{figure}[thb]
\begin{center}
\includegraphics[width=0.70\linewidth]{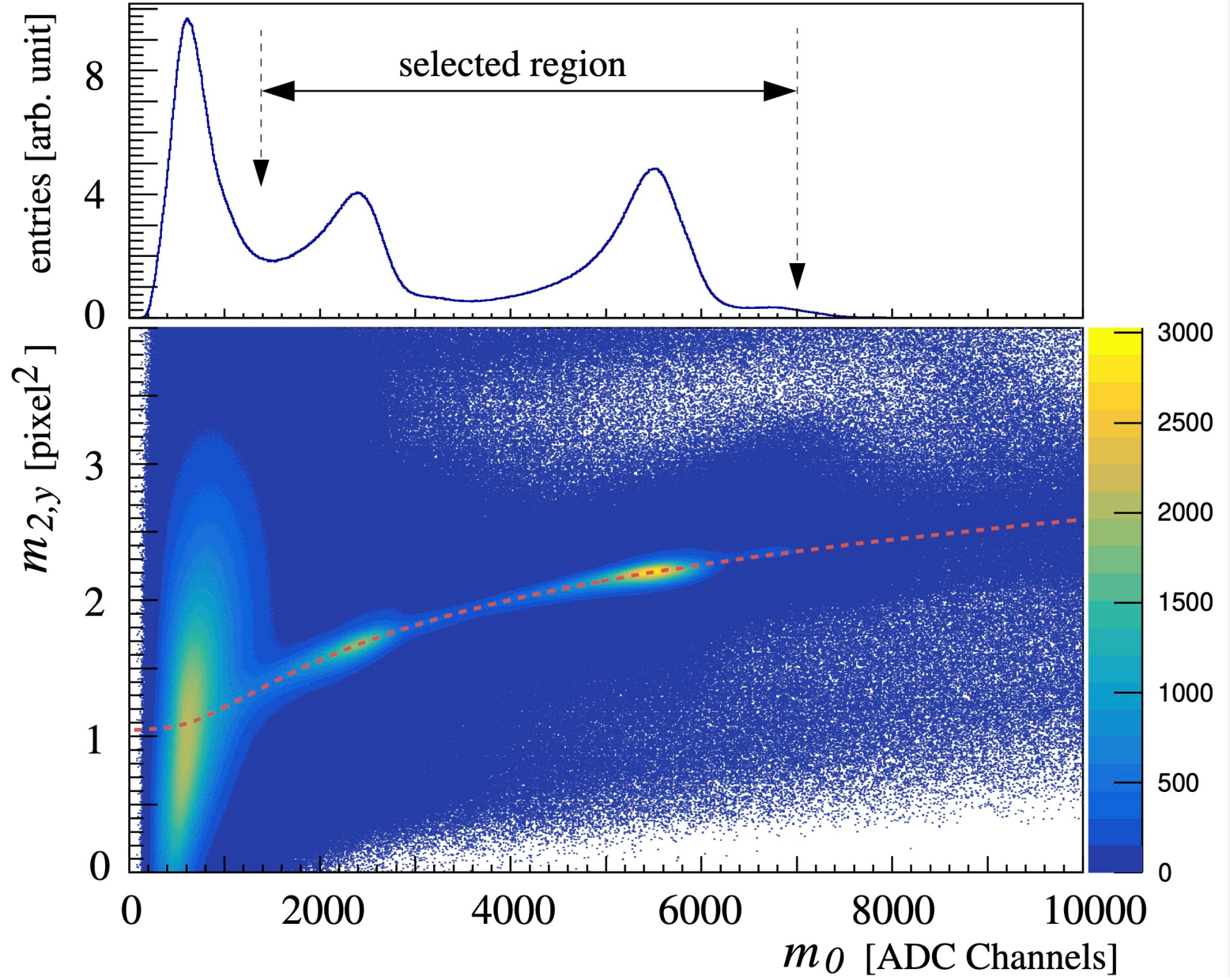}
\caption{
The figure below is a scatter plot of $m_{2,y}$ vs. $m_0$.
Neutron events are expected to lie along the dashed line.
The upper figure shows the $m_0$ distribution of neutron events 
defined within $|\Delta m_{2,y}| < 0.18 $ around the dashed line.
Events in the range of 1400 channels $< m_0 <$ 7000 channels are used for the analysis.
}
\label{fig:nID}
\end{center}
\end{figure}

\section{Event Identification}

The secondary particle generated from the $^{10}\mathrm{B}\,(n, \alpha){}^{7}\mathrm{Li}$ interactions 
loses all their kinetic energy within a few microns in the sensitive region through ionization processes. 
The resulting charges drift toward the electrodes while diffusing in accordance with the internal electric field.
Neutron-induced events are typically confined within a few pixels in our sensor configuration, 
appearing as symmetrical clusters. 
In contrast, clusters originating from background $\gamma$-rays tend to exhibit elongated shapes along the trajectory of a Compton electron. 
These clusters can be distinguished from the neutron events by analysing the characteristics of their shapes (see Fig. \ref{fig:nova}(b)).
The clustering and particle classification method employed in this paper follows previously established approaches \cite{ref:b10int4}.\\

The pixel with the highest signal in the surrounding area is designated as the cluster seed, 
and a $7 \times 7$ pixel region around it is defined as the cluster area. 
By denoting the collected charge in each pixel as $q_i$, 
the characteristics of the cluster are defined as follows:
\begin{equation}
m_0 = \sum_{i=1}^{7\times7} q_i ~, ~~~~ \vec{m}_1 = \frac{1}{m_0} \sum_{i=1}^{7\times7} q_i \vec{r}_i ~, ~~~~ \vec{m}_2 = \frac{1}{m_0} \sum_{i=1}^{7\times7} q_i (\vec{r}_i - \vec{m}_1)^2 ~,
\end{equation}
where $\vec{r}_i = (x_i, y_i)$ is the position of the pixel.

The figure below in the Fig. \ref{fig:nID} is 
a scatter plot of the $y$-component of $\vec{m}_2$ versus $m_0$.
Neutron-induced events are distributed along the dashed line (called it as a neutron line, hereafter).
Events around $m_0 = 2200$ channels originate from $^{7}$Li particles, 
while those around $m_0 = 5500$ channels originate from $\alpha$ particles. 
Events falling within $ |\Delta m_{2,y}| < 0.18 $ pixel$^{2}$ along the neutron line are extracted as neutron events,
where $\Delta m_{2,y}$ represents the deviation from it.
The upper graph in the Fig. \ref{fig:nID} shows the $m_0$ distribution of these extracted neutron events.
The low-energy side includes contributions from $\gamma$-ray background, 
while the high-energy side consists of multiple-count events where multiple event clusters
are present within the $7 \times 7$ pixel cluster region.
To eliminate these contributions, the range of 1400 channels $< m_0 <$ 7000 channels is utilized for the subsequent analysis.\\

\begin{figure}[thb]
\begin{center}
\includegraphics[width=0.81\linewidth]{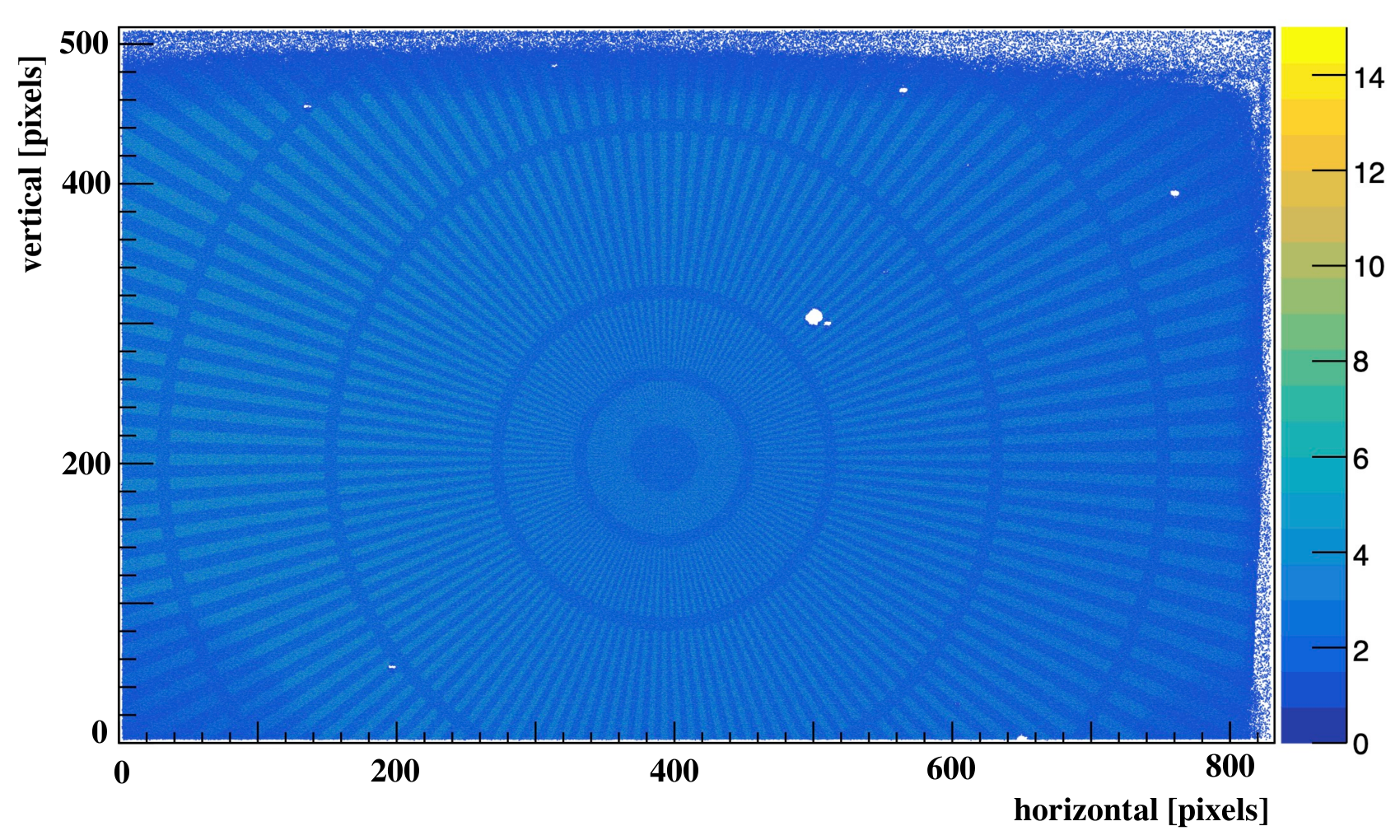}
\caption{
Obtained image of the Siemens Star Chart.
The spokes in the outer ring of the first doughnut shape are clearly distinguishable.
Toward the innermost ring, the chart image gradually becomes unclear due to the effect of beam divergence.
}
\label{fig:hxy}
\end{center}
\end{figure}

Figure \ref{fig:hxy} illustrates the Siemens Star Chart image obtained from the experiment. 
This figure is presented as a two-dimensional histogram with square bins of 0.2 pixels in width, 
with the number of entries per bin indicated by the accompanying color bar. 
Approximately 10 events per bin were recorded. 
The areas with fewer events near the edges of the sensor correlate with regions where the ${}^{10}$B film is thinner.

\section{Imaging Accuracy}
\begin{figure}[thb]
\begin{center}
\includegraphics[width=0.60\linewidth]{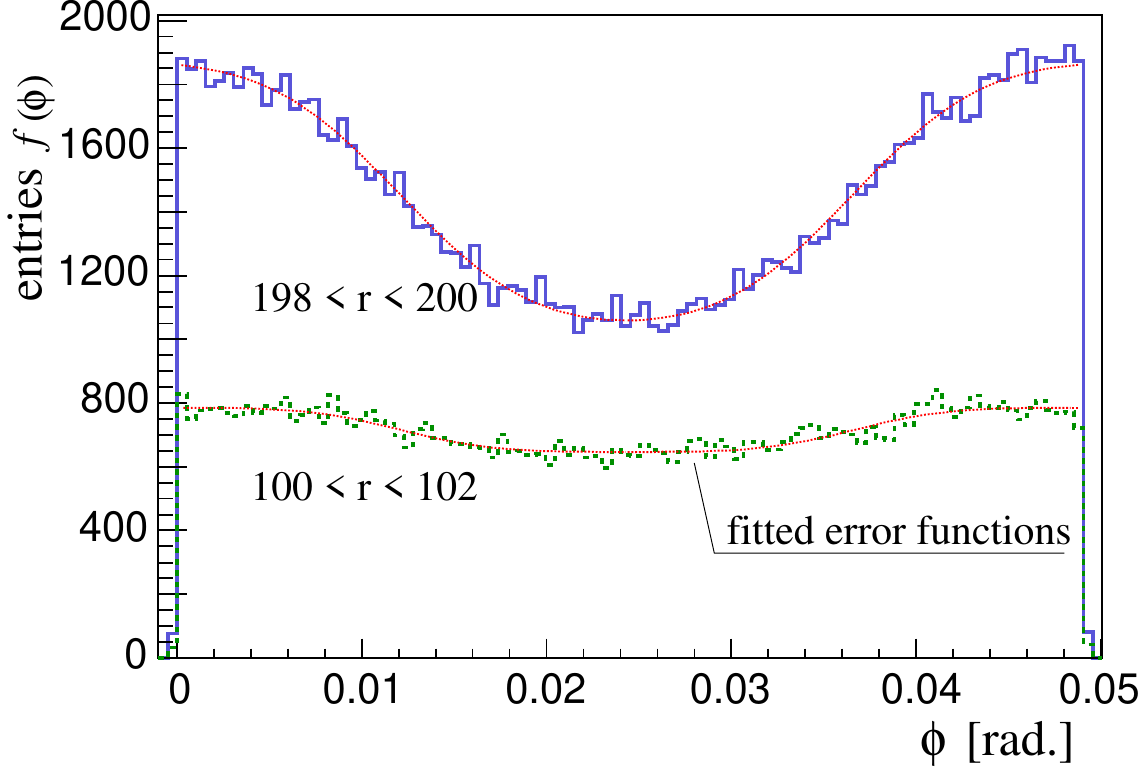}
\caption{
Event distribution of the circular arcs for 100 pixels $< r <$ 120 pixles and 198 pixels $< r <$ 200 pixles.
Dotted lines represent the best fits of the function (\ref{eq:error}).
}
\label{fig:modphi}
\end{center}
\end{figure}

To assess the accuracy of the image, the obtained two-dimensional image is segmented into 128 sets in the angular direction, 
and each set is overlaid within the range of $0 \le \phi < d\phi$.
The sector is further divided into circular arcs with a radial thickness of $dr = 2$ pixels,
and a $\phi$ distribution at radius $r$ is prepared for each arc. 
Figure \ref{fig:modphi} presents the aggregated $\phi$ distributions for the arcs 
corresponding to the ranges 100 pixels $< r <$ 102 pixels and 198 pixels $< r <$ 200 pixels.
In this analysis, the periodic origin is adjusted so that the edges of the Gd pattern align with $\phi = \frac{1}{4}d\phi $ and  $ \phi = \frac{3}{4}d\phi $.\\

The spatial resolution observed from the edge image is evaluated by fitting the following function:
\begin{equation}
f(\phi;r) = a(r) + b(r) \left\{ \mathrm{erfc} \left( \frac{\phi-\frac{1}{4}d\phi}{\sqrt{2}\sigma_{\phi}(r)} \right)  +  \mathrm{erf} \left( \frac{\phi-\frac{3}{4}d\phi}{\sqrt{2}\sigma_{\phi}(r)} \right)  + 2 \right\} ~,
\label{eq:error}
\end{equation}
where $\mathrm{erf}(x)$ and $\mathrm{erfc}(x)$ are the error function and the complementary error function,
and $a(r)$ and $b(r)$ are constants.
$\sigma_{\phi}(r)$ represents the sharpness of the image in the direction perpendicular to the edge, 
corresponding to the rms of the LSF when modeled as a Gaussian.\\

\begin{figure}[thb]
\begin{center}
\includegraphics[width=0.95\linewidth]{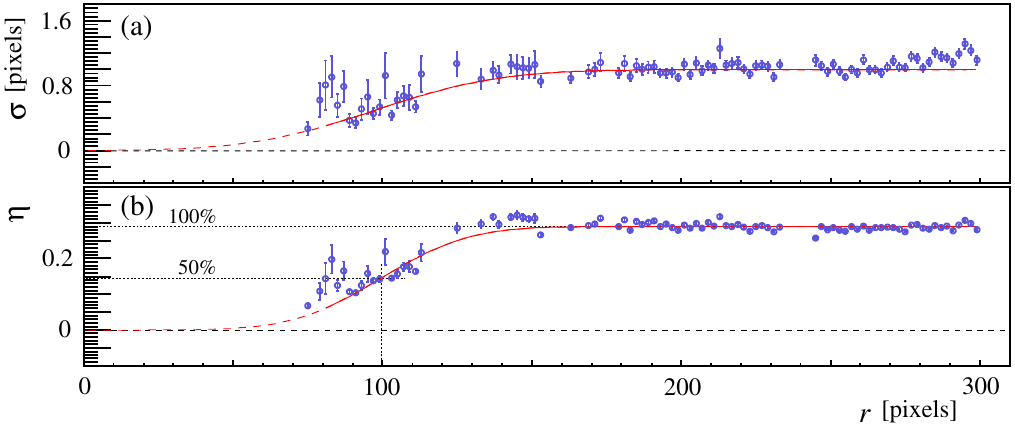}
\caption{
(a) The plot shows the degree of blurriness for each radius $r$.
The dependency, as indicated by the error function, is represented by a curve. 
(b) The plot illustrates the contrast for each radius $r$.
The curve represents the contrast trend fitted with an error function.
The 50\% cutoff structure corresponds to the arc of $r = r_{50}$.
}
\label{fig:reso}
\end{center}
\end{figure}

Figure \ref{fig:reso}(a) displays the estimated rms, $\hat{\sigma} (r)$, where $\hat{\sigma} (r) = r\hat{\sigma}_{\phi}(r)$.
The hat notation indicates the best fit value. 
The spatial resolution is determined to be 
$\sigma_0 = 1.0 $ pixels
by averaging the values in the range of 160 pixels $< r <$ 240 pixels (within the 3rd doughnut).
In the region where the radius is smaller than approximately 120 pixels, 
$\hat{\sigma}(r)$ appears to be underestimated. 
This can be attributed to the inconsistency inherent in the estimation method using the error function fitting.
We assess the degree of inconsistency using one-dimensional Monte Carlo simulations, 
representing the results with a curve,
\begin{equation}
\sigma (r) = \frac{\sigma_0}{2} \left\{ \mathrm{erf} \left( \frac{r-r_{\sigma_{50}}}{p} \right) + 1 \right\} ~,
\end{equation}
here, $r_{\sigma_{50}}$ and $p$ are constants.
The spatial resolution, estimated based on the degree of underestimation characterized by the best-fit value of $r_{\sigma_{50}}$ as
\begin{equation}
\sigma_0 = \frac{2\pi}{128 \times 2} \cdot \frac{\hat{r}_{\sigma_{50}}}{1.87}  ~,
\end{equation}
is 1.3 pixels.
This result is consistent with the previously determined resolution, where $\hat{r}_{\sigma_{50}}$ was 98 pixels.
 
The sharpness of the obtained images is influenced by the inherent spatial resolution of the imager,
the accuracy of the evaluation chart, and the divergence of the beam.
In our setup, the chart's accuracy is sufficiently high to be considered negligible, 
and with the imager's spatial resolution below 5 $\mu$m, its impact is expected to be minimal.
If the image blurriness is primarily attributed to beam divergence,
given that the distance between the chart and the sensor is 6 mm, the estimated beam divergence is 2.7 mrad.
This corresponds to an effective size of the neutron beam at the collimator is approximately 18 mm, and the L/D ratio of 370,
which aligns well with the NOVA design's geometrical L/D ratio of approximately 330.\\

The contrast of the image is evaluated using the following formula,
\begin{equation}
\eta(r) = \frac{\hat{b}(r)}{\hat{a}(r) + \hat{b}(r)} ~,
\end{equation}
where $\hat{a}(r)$ and $\hat{b}(r)$ are the best-fit values of function (\ref{eq:error}).
Figure \ref{fig:reso}(b) shows the contrast distribution as a function of radius $r$.
Points where the estimated contrast exceeded 0.34 or was negative have been excluded.
By averaging the values in the range of 160 pixels $< r <$ 240 pixels,
the maximum contrast is found to be 
$\eta_0 = 0.29$.
This corresponds to about 45\% attenuation in the 5 $\mu$m Gd foil.
The curve represents the trend of the contrast and is fitted using the function
\begin{equation}
\eta (r) = \frac{\eta_0}{2} \left\{ \mathrm{erf} \left( \frac{r-r_{50}}{q} \right) + 1 \right\} ~,
\end{equation}
here, $r_{50}$ and $q$ are constants.
The ratio 
\begin{equation}
\mathrm{MTF}(r) = \frac{\eta(r)}{\eta_0}
\end{equation}
represents the MTF of this imaging system,
reaching 50\% at $r = r_{50}$.
The best fit value of $r_{50}$ was 99 pixels, and
the structure at the 50\% cutoff corresponds to 12 line-space pairs per mm.

\section{Summary}
At J-PARC MLF BL21 (NOVA), image of the Siemens Star Chart was obtained using the ${}^{10}$B-INTPIX4 neutron imager. 
Upon analyzing the image, 
it is determined that the sharpness of the obtained image corresponds to a blurriness of 1.0 pixel $ = 17$ $\mu$m, 
as indicated by the rms of the LSF.
If this blurriness is attributed to beam divergence, it corresponds to an rms beam divergence of 2.7 mrad.
suggesting that the effective size of the neutron beam at the collimator is approximately 18 mm and the L/D ratio of 370.
By configuring the system to capture the entire target plane, 
this device can be effectively used to estimate or monitor the size of a neutron beam, 
even at locations such as neutron production targets that are inaccessible due to high radiation levels.
The structure corresponding to the 50\% cutoff of the MTF is found to be 12 line-space pairs per mm.

\section*{Acknowledgement}
This research was supported by VDEC, The University of Tokyo, 
in collaboration with Cadence Design Systems and NIHON SYNOPSYS G.K. 
It was also funded by JSPS KAKENHI (Grant Nos. 23H00106, 18H04343, 18H01226, 17H05397), 
the TIA Kakehashi program (2024/2023/2022/2021), and the CIQuS project. 
The neutron irradiation test was conducted under the Neutron Scattering Program Advisory Committee of IMSS, KEK (Proposal No. 2024S06).

\end{document}